\documentclass[11pt]{article}
\usepackage{amssymb}
\usepackage{amsmath}
\usepackage{amscd}
\usepackage{latexsym}

\catcode`@=11
\renewcommand{\section}{\@startsection{section}{1}{0pt}{\medskipamount}
{\medskipamount}{\large\bf}}
\numberwithin{equation}{section}
\catcode`@=12

\def\g{\gamma}

\def\de{\delta}

\def\th{\theta}

\def\m{\mu}
\def\n{\nu}

\def\p{\phi}

\def\o{\Omega}
\def\La{\Lambda}

\newcommand{\xh}{\hat{x}}
\newcommand{\yh}{\hat{y}}
\newcommand{\ybh}{\hat{\bar{y}}}
\newcommand{\yb}{\bar{y}}
\newcommand{\ab}{\bar{a}}
\newcommand{\bb}{\bar{b}}
\newcommand{\cb}{\bar{c}}

\newcommand{\C}{\mathbb C}
\newcommand{\R}{\mathbb R}
\newcommand{\Hcal}{{\cal H}}
\newcommand{\Acal}{{\cal A}}

\def\im{\mbox{i}}
\def\N2{$N{=}2$}

\def\diff{\mbox{d}}

\def\sfrac#1#2{{\textstyle\frac{#1}{#2}}}
\def\>{\rangle}
\def\<{\langle}
\def\+{\dagger}
\def\={\ =\ }
\def\und{\qquad\textrm{and}\qquad}
\def\tU{\textrm{U}}

\begin{document}

\begin{titlepage}
\setcounter{page}{0}
\begin{flushright}
hep-th/0603125\\
ITP--UH--05/06\\
\end{flushright}

\vskip 2.0cm

\begin{center}

{\Large\bf Noncommutative Instantons on ${\C}P^n$
}

\vspace{10mm}

{\Large
Tatiana A. Ivanova${}^\+$ \ and \ 
Olaf Lechtenfeld${}^*$ }
\\[5mm]
\noindent ${}^\+${\em Bogoliubov Laboratory of Theoretical Physics, JINR\\
141980 Dubna, Moscow Region, Russia}\\
{Email: ita@thsun1.jinr.ru}
\\[5mm]
\noindent ${}^*${\em Institut f\"ur Theoretische Physik,
Universit\"at Hannover \\
Appelstra\ss{}e 2, 30167 Hannover, Germany }\\
{Email: lechtenf@itp.uni-hannover.de}

\vspace{10mm}

\begin{abstract}
\noindent
We construct explicit solutions of the Hermitian Yang-Mills equations 
on the noncommutative space $\C_\th^n$. In the commutative limit they coincide 
with the standard instantons on $\C P^n$ written in local coordinates. 

\end{abstract}

\end{center}
\end{titlepage}

\section{Introduction and summary}

\noindent
Noncommutative deformations of gauge field theory provide 
a controlled theoretical framework beyond locality~\cite{SWDNS}.
Of particular importance are noncommutative instantons
(see e.g.~\cite{NSNCH, LPTWI} and references therein),
which are BPS~configurations in four dimensions solving
the Yang-Mills self-duality equations.
In the string context, these solutions describe arrangements of 
noncommutative branes (see e.g.~\cite{DHWB} and references therein).

Natural BPS-type equations for gauge fields in more than four
dimensions~\cite{CW, DU} appear in superstring compactification as 
the conditions for the survival of at least one supersymmetry~\cite{GSW88}.
Various solutions to these first-order equations were found
e.g.~in~\cite{FNP, GFKL}, and their noncommutative 
generalizations have been considered e.g.~in~\cite{ncgen, Po}.
For U$(n)$ gauge theory on a K\"ahler manifold these BPS-type equations
specialize to the Hermitian Yang-Mills equations~\cite{DU}.

In this Letter we consider the noncommutative space $\C^n_\th$ and construct 
an explicit $u(n)$-valued solution of the Hermitian Yang-Mills equations. 
In the commutative limit our configuration coincides with the instanton 
solution on $\C P^n$ given in local coordinates on a patch $\C^n$ of $\C P^n$.
We also describe a noncommutative deformation of a local form of the 
Abelian configuration on $\C P^n$. 

\vspace{5mm}

\section{Noncommutative space $\R^{2n}_\th$}

\noindent
Classical field theory on the noncommutative deformation~$\R^{2n}_\th$ of 
the space~$\R^{2n}$ may be realized in a star-product formulation or in an 
operator formalism~\cite{SWDNS}. While the first approach alters the product of
functions on~$\R^{2n}$ the second one turns these functions~$f$ into 
operators~$\hat f$ acting on the $n$-harmonic-oscillator Fock space~$\cal H$. 
The noncommutative space~$\R^{2n}_\th$ may then be defined by
declaring its coordinate functions $\hat x^\mu$ with $\mu =1,\ldots,2n$ to obey
the Heisenberg algebra relations
\begin{equation}
[ \xh^\mu\,,\,\xh^\nu ] \= \im\,\th^{\mu\nu}
\end{equation}
with a constant antisymmetric tensor~$\th^{\mu\nu}$.
The coordinates can be chosen in such a way that the matrix~$(\th^{\m\n})$
will be block-diagonal with non-vanishing components
\begin{equation}\label{tha}
\th^{{2a-1}\ {2a}} \= -\th^{{2a}\ {2a-1}} \ =:\ 
\th^a\quad\mbox{for}\quad a=1,\ldots ,n \ .
\end{equation}
We assume that all $\th^a\ge0$;
the general case does not hide additional complications.
Both approaches are related by the Moyal-Weyl map~\cite{SWDNS}.

{}For the noncommutative version of the complex coordinates
\begin{equation}\label{yyb}
y^a\=x^{2a-1}+\im\,x^{2a} \qquad\textrm{and}\qquad
\yb^{\ab}\=x^{2a-1}-\im\,x^{2a}
\end{equation}
we have
\begin{equation}\label{yhybh}
[\yh^a,\ybh^{\bb} ] \= 2\de^{a\bb}\,\th^a \
=:\ \th^{a\bb} \ge 0\ .
\end{equation}
The Fock space~$\Hcal$ is spanned by the basis states
\begin{equation}
|k_1,k_2,\ldots,k_n\>\=\prod_{a=1}^{n}(2\th^a k_a!)^{-1/2}(\ybh^{a})^{k_a} |
0,\ldots ,0\>
\quad \textrm{for} \quad k_a=0,1,2,\ldots \ ,
\end{equation}
which are connected by the action of creation and annihilation operators
subject to
\begin{equation}
\Bigl[\,\frac{\yh^{a}}{\sqrt{2\th^a}}\ ,\ \frac{\ybh^{\bb}}{\sqrt{2\th^b}}\,
\Bigr] \= \de^{a\bb} \ .
\end{equation}
For simplicity we consider the case $\th^a =\th$ for all~$a$ 
and drop the hats from now on.

\vspace{5mm}

\section{Flat $u(n{+}1)$-connection on $\C^n_\th$}

\noindent
We begin by collecting the coordinates into\footnote{
Here, $\+$ means Hermitian conjugation.}
\begin{equation}
Y\ :=\ \begin{pmatrix}y^1 \\ \vdots \\ y^n \end{pmatrix} \und
Y^\+ \= (\yb^1,\ldots,\yb^n) \ ,
\end{equation}
so that
\begin{equation}
Y^\+ Y \= \yb^a y^a \= \g^2-1-n\th
\end{equation}
with the definition
\begin{equation}
\g\ :=\ \sqrt{x^\m x^\m +1}\= \sqrt{\yb^a y^a +1 + n\th} \ .
\end{equation}
As this is an invertible operator, 
we may also introduce the $n{\times}n$ matrix
\begin{equation}
\La\ :=\ {\bf 1}_n\ -\ Y\frac{1}{\g\,(\g +\sqrt{1{+}n\th})}Y^\+ \ ,
\end{equation}
which obeys
\begin{equation}\label{idntts}
\Lambda\,Y\=Y\,\frac{\sqrt{1{+}n\th}}{\g} \quad,\qquad 
Y^\+\La \=\frac{\sqrt{1{+}n\th}}{\g}\,Y^\+ \und
\La^2\={\bf 1}_n -Y\frac{1}{\g^2}Y^\+\ .
\end{equation}
Since all matrix entries are operators acting in the Fock space~$\Hcal$,
their ordering is essential, in constrast to the commutative case. 
In the present section and the following one, all objects are operator-valued 
in this sense.

Basic for our construction are the $(n{+}1){\times}(n{+}1)$ matrices
\begin{equation}\label{VV+}
V =  \begin{pmatrix}\sqrt{1{+}n\th}\,\g^{-1} & -\g^{-1}Y^\+ \\
                    Y\g^{-1}                 & \La          \end{pmatrix} \und
V^\+=\begin{pmatrix}\sqrt{1{+}n\th}\,\g^{-1} & \g^{-1}Y^\+ \\
                    -Y\g^{-1}                & \La\end{pmatrix} \ .
\end{equation}
With the help of the identities~(\ref{idntts}), one can show that 
\begin{equation}
V^\+V\={\bf 1}_{n+1}\=VV^\+ \quad,\qquad\textrm{i.e.}\quad 
V\in\tU(n{+}1)\ .
\end{equation}
Using $V$, we build a connection one-form
\begin{equation}\label{Acal}
\Acal\= V^\+\diff V \ ,
\end{equation}
which defines the zero curvature
\begin{equation}\label{Fcal}
{\cal F}\=\diff\Acal + \Acal\wedge\Acal \= 
\diff V^\+\wedge\diff V + V^\+\diff V\wedge V^\+\diff V\=0
\end{equation}
on the free module $\C^{n+1}{\otimes}\Hcal$ 
over $\C^n_\th$.

\vspace{5mm}

\section{Nontrivial $u(1)$ and $u(n)$ gauge fields}

\noindent
Let us rewrite $\Acal$ of (\ref{Acal}) in the block form
\begin{equation}\label{Ablock}
\Acal \=\begin{pmatrix} a  & -\p^\+ \\
                        \p & A      \end{pmatrix} 
\qquad\textrm{with}\qquad a\in u(1) \quad\textrm{and}\quad A\in u(n) \ ,
\end{equation}
Clearly, $\p$ is an $n{\times}1$ matrix and $\p^\+$ its hermitian conjugate. 
{}From the definition (\ref{Acal}) we find that
\begin{align}
\label{a}
a &\= \g\, \diff\g^{-1} + \g^{-1}Y^\+(\diff Y)\g^{-1}\ ,\\[8pt]
\label{A}
A &\= Y\g^{-1}(\diff\g^{-1}) Y^\+ +Y\g^{-2}\diff Y^\+{+}\La\,\diff\La\ ,\\[8pt]
\p &\= \La\,(\diff Y)\, \g^{-1} \= \bigl(
\diff Y-Y(\g^2+\g\,\sqrt{1{+}n\th})^{-1}\,Y^\+\diff Y\bigr)\g^{-1}\ ,\\[8pt]
\label{p+}
\p^{\+} &\= \g^{-1}(\diff Y^\+)\La \= \g^{-1}
\bigl(\diff Y^\+ -(\diff Y^\+)Y(\g^2+\g\,\sqrt{1{+}n\th})^{-1}\,Y^\+ \bigr) \ .
\end{align}
Introducing the components $\p^a$ of the column $\p = (\p^a)$,
the last two equations read
\begin{align}
\p^a &\= \bigl(\diff y^a - y^a\,(\g^2 +\g\,\sqrt{1{+}n\th} )^{-1}\, 
\de_{\bb c}\, \yb^{\bb}\, \diff y^c \bigr) \, \g^{-1}\ , \\[8pt]
{\bar{\p}}^{\ab} &\= \g^{-1}\bigl(\diff \yb^{\ab} - \diff\yb^{\cb}\, 
\de_{b\cb}\, y^b\, (\g^2 +\g\,\sqrt{1{+}n\th} )^{-1}\,\yb^{\ab}\bigr) \ .
\end{align}
The (1,0)-forms $\p^a$ and the (0,1)-forms ${\bar{\p}}^{\bb}$ constitute 
a basis for the forms of type (1,0) and~(0,1), respectively.

Substituting (\ref{Ablock}) into (\ref{Fcal}), we obtain
\begin{align}
\label{Fu1}
F_{u(1)} &\ :=\ \diff a + a\wedge a \,\= \,\p^\+\wedge\p 
\= \de_{\ab b}\,\bar{\p}^{\ab}\wedge\p^b 
\= \bar{\p}^1\wedge\p^1 + \ldots + \bar{\p}^n\wedge\p^n \ , \\[8pt] 
\label{Fun}
F_{u(n)} &\ :=\ \diff A {+} A\wedge A \= \p\wedge\p^\+ 
\= (\p^a\wedge\bar{\p}^{\bb}) 
\= \begin{pmatrix} 
\p^1\wedge\bar{\p}^{\bar{1}} & \cdots & \p^1\wedge\bar{\p}^{\bar{n}} \\
\vdots                       & \ddots & \vdots                       \\
\p^n\wedge\bar{\p}^{\bar{1}} & \cdots & \p^n\wedge\bar{\p}^{\bar{n}} 
\end{pmatrix}
\end{align}
as well as
\begin{equation}
0 \= \diff\p + \p\wedge a + A\wedge\p \und
0\= \diff\p^\+ + a\wedge \p^\+ + \p^\+\wedge A \ .
\end{equation}
{}From (\ref{Fu1}) and (\ref{Fun}) one sees that the gauge fields $F_{u(1)}$ 
and $F_{u(n)}$ have vanishing (2,0) and (0,2) components, i.e.~they are
of type~(1,1). Moreover, (\ref{Fun}) expresses $F_{u(n)}$ in the basis 
$\{\p^a\wedge\bar{\p}^{\bb}\}$ of (1,1)-forms as
\begin{equation}
F_{u(n)} \= F_{a\bb}\,\p^a\wedge \bar{\p}^{\bb} \qquad\Longrightarrow\qquad
F_{ab} \= 0 \= F_{\ab\bb} \und F_{a\bb} \= e_{ab} \= -F_{\bb a} \ ,
\end{equation}
where the basis matrix $e_{ab}$ has a unit entry in the $(ab)$ position 
and is zero elsewhere.

It is apparent that the operator-valued components of the $u(n)$-valued 
gauge field $F_{u(n)}$ satisfy the Hermitian Yang-Mills equations\footnote{
Their general form for the structure group U$(k)$ reads 
$F_{ab}{=}0{=}F_{\ab\bb}\ ,\ F_{1\bar{1}}+\ldots +F_{n\bar{n}}{=}\tau{\bf 1}_k$,
where $\tau$ is a constant.}
\begin{equation}
F_{ab} \= 0 \= F_{\ab\bb} \und
F_{1\bar{1}}+\ldots +F_{n\bar{n}} \= {\bf 1}_n\ .
\end{equation}
In the commutative case these equations are the conditions of stability 
for a holomorphic vector bundle over $\C P^n$ with finite characteristic 
classes~\cite{DU}.
In the star-product formulation obtained by the inverse Moyal-Weyl transform, 
the gauge field~(\ref{Fun}) describes a smooth Moyal deformation of the 
instanton-type gauge field configuration given in local coordinates on a patch
$\C^n$ of~$\C P^n$.
This is why we call the configuration (\ref{A}) and~(\ref{Fun}) the 
`noncommutative U$(n)$ instanton on~$\C P^n$'.
Likewise, the Abelian field strength~(\ref{Fu1}) with components 
$f_{a\bb}:=-\delta_{a\bb}$ satisfies the Hermitian Maxwell equations
\begin{equation}
f_{ab} \= 0 \= f_{\ab\bb} \und
f_{1\bar{1}}+\ldots +f_{n\bar{n}} \= -n \ ,
\end{equation}
whence the configuration (\ref{a}) and (\ref{Fu1}) is the 
`noncommutative U$(1)$ instanton on~$\C P^n$'.

\vspace{5mm}

\section{Commutative limit}

\noindent
In the commutative limit, $\th\to0$, the gauge potential $A$ defining 
$F_{u(n)}$ coincides with the instanton-type canonical connection on~$\C P^n$, 
which is described as follows~\cite{NR}. Consider the group U$(n{+}1)$, 
its Grassmannian subset $\C P^n =\tU(n{+}1)/\tU(1){\times}\tU(n)$ 
and the fibration
\begin{equation}\label{bundle}
\begin{CD}
\tU(n{+}1)@>{\tU(1)\times\tU(n)}>> \C P^n
\end{CD}
\end{equation}
with fibres $\tU(1){\times}\tU(n)$. For $g\in\tU(n{+}1)$ the canonical one-form
$\o = g^\+\diff g$ on U$(n{+}1)$ takes values in the Lie algebra $u(n{+}1)$
and satisfies the Maurer-Cartan equation
\begin{equation}\label{MC}
\diff \o\ +\ \o\wedge\o \=0\ .
\end{equation}
The matrix $V$ from (\ref{VV+}) defines a local section of the bundle 
(\ref{bundle}) over a patch $\C^n\subset\C P^n$, viz.~the embedding of $\C P^n$
into U$(n{+}1)$. For such an embedding the one-form $\o$ coincides with the 
flat connection~$\Acal$ given by~(\ref{Acal}). It follows that (\ref{Fcal})
is the Maurer-Cartan equation~(\ref{MC}) reduced to $\C^n\subset\C P^n$, 
and the block form~(\ref{Ablock}) results from the splitting of $\Acal$ into
components $\p$ and $\p^\+$ tangent\footnote{
They are basis one-forms on $\C P^n$ taking values in the complexified 
tangent bundle of $\C P^n$.}
to $\C P^n$ and into one-forms $a$ and $A$ on $\C P^n$ with values in the 
tangent space $u(1){\oplus}u(n)$ to the fibre U$(1){\times}$U$(n)$
of the bundle~(\ref{bundle}). 

By construction, the one-form $A$ from~(\ref{A}) is the canonical connection 
on the Stiefel bundle 
\begin{equation}\label{st1}
\begin{CD}
\tU(n{+}1)/\tU(1) @>{\tU(n)}>> \C P^n
\end{CD}
\end{equation} 
given by~\cite{NR}
$$
A\={\cal S}^\+ \diff {\cal S}\ ,
$$
where $\cal S$ is an $(n{+}1){\times}n$ matrix-valued section of the bundle (\ref{st1})
such that ${\cal S}^\+{\cal S}={\bf 1}_n$. In our case it is chosen as
\begin{equation}
{\cal S}\=\begin{pmatrix} -\g^{-1}Y^\+ \\ \La \end{pmatrix}\ ,
\end{equation}
i.e.~as the $(n{+}1){\times}n$-part of the matrix $V$ from (\ref{VV+}). 
Similarly, the one-form~$a$ from~(\ref{a}) in the commutative limit 
coincides with the canonical Abelian connection 
\begin{equation}
a\={s}^\+\diff {s}
\end{equation}
on another Stiefel bundle:
\begin{equation}\label{st2}
\begin{CD}
S^{2n+1}\=\tU(n{+}1)/\tU(n) @>{\tU(1)}>> \C P^n\ .
\end{CD}
\end{equation}  
In our case,  
$s=\left( \begin{smallmatrix} 1 \\ Y \end{smallmatrix} \right)\g^{-1}$ 
is the $(n{+}1){\times}1$ matrix complementing $\cal S$ inside the matrix $V$.
Moreover, the Abelian gauge field 
$F_{u(1)}= -\de_{a\bb}\ \p^a\wedge \p^{\bb}$
is proportional to the two-form 
\begin{equation}
\omega \=\sfrac{\im}{2}\, \de_{a\bb}\ \p^a\wedge \p^{\bb}\ ,
\end{equation}
which is the canonical K\"ahler two-form on $\C P^n$. 

\vspace{10mm}

\noindent
{\bf Acknowledgements}

\medskip

\noindent
T.A.I.~acknowledges the Heisenberg-Landau program and RFBR 
(grant 06-01-00627-a) for partial support and the Institut f\"ur 
Theoretische Physik der Universit\"at Hannover for its hospitality. 
The work of O.L. was partially supported by the Deutsche
Forschungsgemeinschaft (DFG).

\end{document}